\documentclass[a4paper,conference]{IEEEtran}
\usepackage{cite}
\ifCLASSINFOpdf
  \usepackage[pdftex]{graphicx}
\else
  \usepackage[dvips]{graphicx}
\fi
\usepackage[T1]{fontenc}
\usepackage{amsmath}
\usepackage{array}
\usepackage{theorem}
\usepackage[caption=false,font=footnotesize]{subfig}
\usepackage{url}
\usepackage{acronym}
\usepackage{color}

\acrodef{LDPC}{low-density parity-check}
\acrodef{MDPC}{moderate-density parity-check}
\acrodef{QC}{quasi-cyclic}
\acrodef{QC-LDPC}{quasi-cyclic low-density parity-check}
\acrodef{QC-MDPC}{quasi-cyclic moderate-density parity-check}
\acrodef{RSA}{Rivest, Shamir, Adleman}
\acrodef{BF}{bit flipping}
\acrodef{SPA}{sum product algorithm}
\acrodef{RDF}{random difference families}
\acrodef{ISD}{information set decoding}
\acrodef{KRA}{key recovery attack}
\acrodefplural{KRA}[KRAs]{key recovery attacks}
\acrodef{DA}{decoding attack}
\acrodefplural{DA}[DAs]{decoding attacks}
\acrodef{WF}{work factor}
\acrodef{BER}{bit error rate}
\acrodef{CER}{codeword error rate}
\acrodef{BSC}{binary symmetric channel}
\acrodef{BPSK}{binary phase shift keying}
\acrodef{$2$-PAM}{binary pulse amplitude modulation}
\acrodef{AWGN}{additive white Gaussian noise}
\acrodef{LLR}{log-likelihood ratio}
\acrodefplural{LLR}[LLRs]{log-likelihood ratios}
\acrodef{SPA}{sum-product algorithm}
\acrodef{DFR}{decoding failure rate}

\begin{document}

\title{Soft McEliece: MDPC code-based McEliece cryptosystems with very compact keys through real-valued intentional errors}

\author{\IEEEauthorblockN{Marco Baldi, Paolo Santini, Franco Chiaraluce,\\}
\IEEEauthorblockA{DII, Universit\`a Politecnica delle Marche,\\
Ancona, Italy\\
Email: \{m.baldi, f.chiaraluce\}@univpm.it, S1069914@studenti.univpm.it}}

\maketitle

\begin{abstract}
We propose to use real-valued errors instead of classical bit flipping intentional errors in the 
McEliece cryptosystem based on \ac{MDPC} codes.
This allows to exploit the error correcting capability of these codes to the utmost, by using soft-decision
iterative decoding algorithms instead of hard-decision bit flipping decoders.
However, soft reliability values resulting from the use of real-valued noise can also be exploited by attackers.
We devise new attack procedures aimed at this, and compute the relevant work factors and security levels.
We show that, for a fixed security level, these new systems achieve the shortest public key sizes ever reached, 
with a reduction up to $25 \%$ with respect to previous proposals.
\end{abstract}

\begin{IEEEkeywords}
Compact keys, LDPC codes, McEliece cryptosystem, MDPC codes, real-valued intentional errors.
\end{IEEEkeywords}

\section{Introduction}
\label{sec:Intro}
\let\thefootnote\relax\footnotetext{This work was supported in part by the MIUR project ``ESCAPADE''
(Grant RBFR105NLC) under the ``FIRB - Futuro in Ricerca 2010'' funding program.}

Quantum computers are able to penetrate hard cryptographic targets, like cryptosystems
based on integer factorization and discrete logarithms, and code-based cryptosystems are among
the most promising solutions able to resist quantum computer-based attacks.
The McEliece cryptosystem \cite{McEliece1978} is the best known code-based asymmetric cryptosystem.
In its original formulation based on Goppa codes, it achieves very fast encryption and decryption but
has very large public keys, which is a major drawback.
According to \cite{Bernstein2008}, for achieving $80$-bit security with the Goppa code-based cryptosystem
we need $460647$-bit public keys.
A known way to reduce the public key size is to replace Goppa codes with other families of codes,
although this may expose the system to security flaws.
A recent line of research has been focused on the use of \ac{QC-LDPC} and \ac{QC-MDPC} codes 
in this context, showing that practical systems with compact keys can be designed while preserving 
the system security \cite{Baldi2008, Baldi2012, Misoczki2012, Baldi2013, Baldi2013a, Biasi2014, VonMaurich2015}.

This has been achieved by keeping the same structure of the original McEliece cryptosystem, in
which binary intentional errors are used during encryption.
Therefore, hard-decision decoders like the bit flipping iterative decoder \cite{Gallager1963}
are commonly used in these systems.
Message-passing decoders can also be used, but without the availability of soft reliability values 
concerning the ciphertext bits.
This makes such decoders work in suboptimal conditions, that penalize their performance.
In fact, \ac{LDPC} codes achieve the best performance under soft-decision message-passing 
decoding when soft reliability information is available \cite{Richardson2001}.
For such a reason, in this paper we propose to use real-valued noise samples as an
intentional impairment during encryption.
This allows to exploit powerful
soft-decision decoders to improve the error correcting capability of the legitimate receiver.
On the other hand, soft reliability information can be exploited by attackers as well.
Therefore, we develop an updated security analysis taking this fact into account.
Our results show that this approach allows to achieve public key size reductions up to
$25 \%$ with respect to the previously known best solutions.

The paper is organized as follows:
in Section \ref{sec:System} we recall the original \ac{QC-LDPC} and \ac{QC-MDPC} code-based McEliece cryptosystems,
in Section \ref{sec:NewSystem} we introduce a new \ac{QC-MDPC} code-based variant exploiting real-valued intentional noise,
in Section \ref{sec:SecLevel} we assess security of the new system,
in Section \ref{sec:Examples} we provide some design examples and
in Section \ref{sec:Conclusion} we draw some conclusive remarks.

\section{\ac{QC-LDPC} and \ac{QC-MDPC} code-based McEliece cryptosystems}
\label{sec:System}

The original \ac{QC-LDPC} and \ac{QC-MDPC} code-based McEliece cryptosystems exploit codes having rate $R = \frac{n_0-1}{n_0}$,
where $n_0$ is a small integer (\textit{e.g.}, $n_0 = 2,3,4$), redundancy $r$, length $n = n_0 \cdot r$ and dimension 
$k = (n_0 - 1) \cdot r$.
The secret code is defined through a sparse parity-check matrix $\mathbf{H}$ having the following form \cite{Baldi2011a, Baldi2012}:
\begin{equation}
\mathbf{H} = \left[ \mathbf{H}_{0} | \mathbf{H}_{1} | \ldots |\mathbf{H}_{n_0-1} \right],
\label{eq:HCircRow}
\end{equation}
where each block $\mathbf{H}_{i}$ is a circulant matrix with size $r \times r$.
It has been recently shown that odd values of $r$ must be chosen to avoid some possible weaknesses \cite{Londahl2015}.
In most instances of these systems appeared in previous literature, the matrix $\mathbf{H}$ is regular, although it has been shown in \cite{Baldi2013a}
that using irregular matrices may bring some reduction in the public key size.
Thus, for the sake of comparison, we focus on regular parity-check matrices, having all the columns with weight $d_v$ and all the rows with weight $d_c = n_0 \cdot d_v$.
When $d_v$ is much smaller than $r$, we say that the code is an \ac{LDPC} code.
\ac{MDPC} codes are a special class of \ac{LDPC} codes, characterized by moderately small values of $d_v$ (in the order of several tenths).

\subsection{Key generation}

The private key is formed by the secret parity-check matrix $\mathbf{H}$ and two other non-singular matrices: a $k \times k$ scrambling matrix $\mathbf{S}$ 
and an $n \times n$ transformation matrix $\mathbf{Q}$.
The latter is defined as a sparse matrix with average row and column weight $m \ge 1$ ($m$ is not necessarily an integer, since $\mathbf{Q}$ can be irregular).  
Both $\mathbf{S}$ and $\mathbf{Q}$ have \ac{QC} form, that is, they consist of circulant sub-matrices with size $r \times r$.
When \ac{QC-MDPC} codes are used, $\mathbf{Q}$ boils down to an $n \times n$ permutation matrix, and we have $m=1$.
In this case, as done in \cite{Misoczki2012}, the secret permutation can even be avoided and $\mathbf{Q}$ eliminated (\textit{i.e.}, $\mathbf{Q} = \mathbf{I}_{n \times n}$, the $n \times n$ identity matrix).

The public key is obtained as $\mathbf{G}' = \mathbf{S}^{-1} \cdot \mathbf{G} \cdot{\mathbf{Q}^{-1}}$, where $\mathbf{G}$ is a systematic generator matrix obtained from $\mathbf{H}$.
The role of $\mathbf{S}$ is to make the public generator matrix non-systematic.
However, if we consider a CCA2 secure conversion of the system \cite{Bernstein2008}, $\mathbf{G}'$ can be in systematic form, therefore $\mathbf{S}$ can be eliminated (\textit{i.e.}, $\mathbf{S} = \mathbf{I}_{k \times k}$).
With $\mathbf{G}'$ in systematic form, and exploiting its \ac{QC} structure, the public key size is $(n_0 - 1) \cdot r$ bits.
Such a size is considerably smaller than for classical Goppa code-based instances with the same security level.

It follows from the public key definition that the public code admits a parity-check matrix in the form $\mathbf{H'} = \mathbf{H \cdot Q}^T$, that is sparse.
Since both $\mathbf{H}$ and $\mathbf{Q}$ are indeed very sparse, $\mathbf{H'}$ is very likely the sparsest parity-check matrix of the public code.
For this reason, it can be the target of a key recovery attack, as we will see in Section \ref{subsec:KRA}.

\subsection{Encryption}

In order to encrypt her message, Alice gets Bob's public key $\mathbf{G}'$, divides her message into $k$-bit vectors and, 
for each of them, generates a random intentional error vector $\mathbf{e}$ with weight $t$.
Finally, she encrypts $\mathbf{u}$ into $\mathbf{x}$ as follows:
\begin{equation}
\mathbf{x} = \mathbf{u} \cdot \mathbf{G}' \oplus \mathbf{e} = \mathbf{c'} \oplus \mathbf{e},
\label{eq:Encryption}
\end{equation}
where $\oplus$ denotes modulo-$2$ addition.
In fact, in all existing McEliece cryptosystems, independently of the family of codes used, the intentional errors are bit flipping errors.
This means that $\mathbf{e}$ is a binary vector, and the bits of the codeword $\mathbf{c'}$ which are at positions corresponding to 
the support of $\mathbf{e}$ are flipped.

\subsection{Decryption}
\label{subsec:Decryption}

In order to perform decryption, Bob first inverts the secret transformation (if used):
\begin{equation}
\mathbf{x}' = \mathbf{x} \cdot \mathbf{Q} = \mathbf{u} \cdot \mathbf{S}^{-1} \cdot \mathbf{G} \oplus \mathbf{e} \cdot \mathbf{Q} = \mathbf{c} \oplus \mathbf{e} \cdot \mathbf{Q}.
\label{eq:xprime}
\end{equation}
This way, he gets the codeword $\mathbf{c}$ belonging to the private code, corrupted by the error vector $\mathbf{e'} = \mathbf{e} \cdot \mathbf{Q}$.
Due to the structure of $\mathbf{Q}$, $\mathbf{e'}$ is a binary vector with weight $\leq t'=t m$.
When $m=1$, as in the case of \ac{QC-MDPC} code-based systems, $\mathbf{Q}$ is a permutation or an identity matrix, hence $t' = t$.
Then, Bob performs \ac{LDPC} decoding to correct all the errors and easily obtains $\mathbf{u} \cdot \mathbf{S}^{-1}$ owing to systematic encoding.
The message $\mathbf{u}$ is then recovered through multiplication by $\mathbf{S}$.

\section{Exploiting real-valued intentional noise}
\label{sec:NewSystem}

Let us consider a new system in which we use a real-valued intentional noise vector $\mathbf{w}$ in place of the binary intentional error vector $\mathbf{e}$.
In other terms, $\mathbf{w}$ is a $1 \times n$ vector containing real-valued noise samples affecting the codeword $\mathbf{c'}$ during encryption.
This way, the ciphertext $\mathbf{x}$ is no longer a binary vector, and becomes a real-valued vector as well.

From the practical standpoint, real numbers are always represented through finite precision variables.
So, we suppose to use $q$-bit variables to represent the entries of $\mathbf{c'}$, $\mathbf{w}$ and $\mathbf{x}$.
The effects of the finite precision representation of real numbers can be made negligible by choosing a suitably large value of $q$.
Obviously, the ciphertext length is increased by a factor $q$ with respect to the classical systems, and this may seem an important drawback. 
However, as we will see next, exploiting real-valued intentional noise allows to achieve significant reductions in the public key size, 
which is the most important drawback of McEliece-type cryptosystems.
Moreover, the intentional noise vector can also be exploited to carry information, as first proposed in \cite{Riek1990}.
This means that we could encode part of the secret message into the intentional noise vector, thus reducing the ciphertext expansion.
Such a possibility, however, is left for future investigation. 

For the sake of simplicity, in the following we describe the main procedures of the new system by focusing on the \ac{QC-MDPC} code-based variant described 
in Section \ref{sec:System} with CCA2 secure conversion, using $\mathbf{S} = \mathbf{I}_{k \times k}$ and $\mathbf{Q} = \mathbf{I}_{n \times n}$.
The extension to more general \ac{QC-LDPC} and \ac{QC-MDPC} code-based schemes is also left for future works.

\subsection{Key generation}

As in the original system, the private key is formed by an $r \times n$ secret parity-check matrix $\mathbf{H}$ in the form \eqref{eq:HCircRow}, 
from which a $k \times n$ generator matrix $\mathbf{G}$ is obtained, in systematic form.
Since $\mathbf{S} = \mathbf{I}_{k \times k}$ and $\mathbf{Q} = \mathbf{I}_{n \times n}$, the public key is $\mathbf{G}' = \mathbf{G}$,
with size $(n_0 - 1) \cdot r$ bits.

\subsection{Encryption}

There are several possibilities to extend the encryption map \eqref{eq:Encryption} to the case in which we use a real-valued intentional noise.
Among these, we choose an encryption map that allows to exploit some \ac{LDPC} coding theory concepts which are well known in the literature.
In fact, a huge amount of research works have been devoted to the design and optimization of \ac{LDPC} coded transmission schemes with antipodal signals (\textit{e.g.},
\ac{$2$-PAM}) over \ac{AWGN} channels.
Such results can be reused in the context under consideration if decryption is performed on a vector which looks like a modulated LDPC codeword with symbols $1$ in place 
of bits $1$ and symbols $-1$ in place of bits $0$, corrupted by \ac{AWGN}.
For this purpose, let us consider the following encryption map
\begin{equation}
\mathbf{x} = 2\left(\mathbf{u} \cdot \mathbf{G}' \right) - \mathbf{1} + \mathbf{w} = 2 \mathbf{c'} - \mathbf{1} + \mathbf{w},
\label{eq:SoftEncryptionGeneral}
\end{equation}
where $\mathbf{1}$ is the $1 \times n$ all-one vector.
The statistical properties of $\mathbf{w}$ can obviously be fixed a priori.
We suppose that $\mathbf{w}$ is filled with the samples of a Gaussian variable with mean $0$ and standard deviation $\sigma$. 
As we will see in Section \ref{subsec:NewDA}, the generation of the vector $\mathbf{w}$ may require more than one attempt, since some results need to be discarded for security reasons.

\subsection{Decryption}

Since $\mathbf{G}' = \mathbf{G}$, the ciphertext $\mathbf{x}$ received by Bob coincides with a \ac{$2$-PAM} modulated version of a codeword $\mathbf{c'} = \mathbf{c}$ belonging to the private code, corrupted by the intentional noise vector $\mathbf{w}$, that contains \ac{AWGN} samples.
As in Section \ref{subsec:Decryption}, Bob then performs \ac{LDPC} decoding to correct all the errors and recovers $\mathbf{u}$ owing to systematic encoding.

Differently from the original system, in this new system Bob can exploit the optimal performance of message-passing \ac{LDPC} decoding algorithms working on the soft reliability values associated to the received samples.
One of the best algorithms of this type is the \ac{SPA} with \acp{LLR}, that we consider in the following.
In order to perform decoding through the \ac{LLR}-\ac{SPA}, Bob needs to compute the \ac{LLR} of each codeword bit, defined as
\begin{equation}
\Lambda(x_i) = \log \left[ \frac{p(x_i|c_i=1)}{p(x_i|c_i=0)} \right],
\label{eq:LLRdef}
\end{equation}
where $x_i$ is the symbol corresponding to the codeword bit $c_i$ impaired with noise, and $p(x_i|c_i)$ is the probability density function (p.d.f.) of $x_i$ conditioned on $c_i$.
By taking into account that we have \ac{$2$-PAM} signals impaired with \ac{AWGN}, through simple calculations \eqref{eq:LLRdef} can be rewritten as
\begin{equation}
\Lambda(x_i) = \frac{2x_i}{\sigma^2},
\label{eq:LLRvalue}
\end{equation}
where $\sigma$ is the \ac{AWGN} standard deviation.

\section{Security assessment}
\label{sec:SecLevel}

In this section we assess the security of the proposed cryptosystem by considering both classical attacks
and newly developed attacks aimed at exploiting the real-valued intentional noise.
There are two main types of attacks which may be mounted against these systems:
\acp{DA} and \acp{KRA}.
While the former are aimed at decrypting one or more ciphertexts without knowing the private key, the latter 
aim at recovering the private key from the public key.
The soft reliability information about the ciphertext bits that is available in the proposed system may facilitate 
\acp{DA}, thus helping an attacker to decrypt an intercepted ciphertext.
For this reason, we consider classical \acp{DA}, but also devise new \acp{DA} exploiting soft reliability information.

\subsection{Classical decoding attacks}
\label{subsec:ClassicalDA}

In a \ac{DA}, the adversary intercepts a ciphertext $\mathbf{x}$ and aims at correcting all the intentional errors
added during encryption.
If this succeeds, he can then invert the encoding map and recover the cleartext.
The most dangerous \acp{DA} against the original \ac{LDPC} and \ac{MDPC} code-based cryptosystems are those
exploiting \ac{ISD} algorithms.
These techniques stem from a family of probabilistic algorithms aimed at finding low weight codewords in general
linear block codes, first introduced by Leon \cite{Leon1988} and Stern \cite{Stern1989}.
Indeed, it can be shown that finding the binary error vector $\mathbf{e}$ that has been used in \eqref{eq:Encryption} 
to obtain the ciphertext is very similar to searching for a low weight codeword in an extended version of the public code.

These algorithms have known great advances in recent years \cite{Peters2010, May2011, Bernstein2011, Becker2012}.
Today, one of the best known algorithms to search for low weight codewords in a linear block code is that
introduced in \cite{Becker2012}. Its work factor in the finite code length regime has been computed in closed
form in \cite[Appendix B]{Misoczki2012}.
For a code with length $n$ and dimension $k$ in which a (single) codeword with weight $w$ is searched,
we define this quantity as $WF_{\mathrm{BJMM}}(n,k,w)$, representing the number of elementary operations
which are needed to successfully complete the algorithm execution, on average.

The \ac{QC} nature of the codes we consider facilitates such a task, since each block-wise cyclically shifted 
version of a ciphertext is still a valid ciphertext.
Therefore, an attacker could consider all the \ac{QC} shifts of an intercepted ciphertext, and search for one 
among as many shifted versions of the error vector.
For codes in the form \eqref{eq:HCircRow}, the number of possible \ac{QC} shifts of a ciphertext is $r$, and
the corresponding advantage in terms of the algorithm complexity is in the order of $\sqrt{r}$ \cite{Sendrier2011}.
Therefore, the work factor of decoding attacks against the original systems can be computed as
\begin{equation}
WF_{\mathrm{DA}}^{(1)}(t) = \frac{1}{\sqrt{r}} WF_{\mathrm{BJMM}}(n,k,t),
\end{equation}
where $t$ is the weight of the vector $\mathbf{e}$ used during encryption.
An attack of this kind can also be mounted against the new cryptosystem by applying hard-decision, 
discarding the soft reliability information and trying to correct the bit errors induced by the intentional noise.

\subsection{Soft reliability information-aided decoding attacks}
\label{subsec:NewDA}

According to the proposed approach, both Bob and Eve may take advantage of the soft reliability information
about each bit of any ciphertext.
This facilitates Bob, which may exploit powerful iterative soft-decision decoding algorithms for \ac{LDPC} and \ac{MDPC} codes.
However, the same information can also be exploited by Eve to mount a decoding attack.

\begin{figure}[!t]
\begin{centering}
\includegraphics[width=80mm,keepaspectratio]{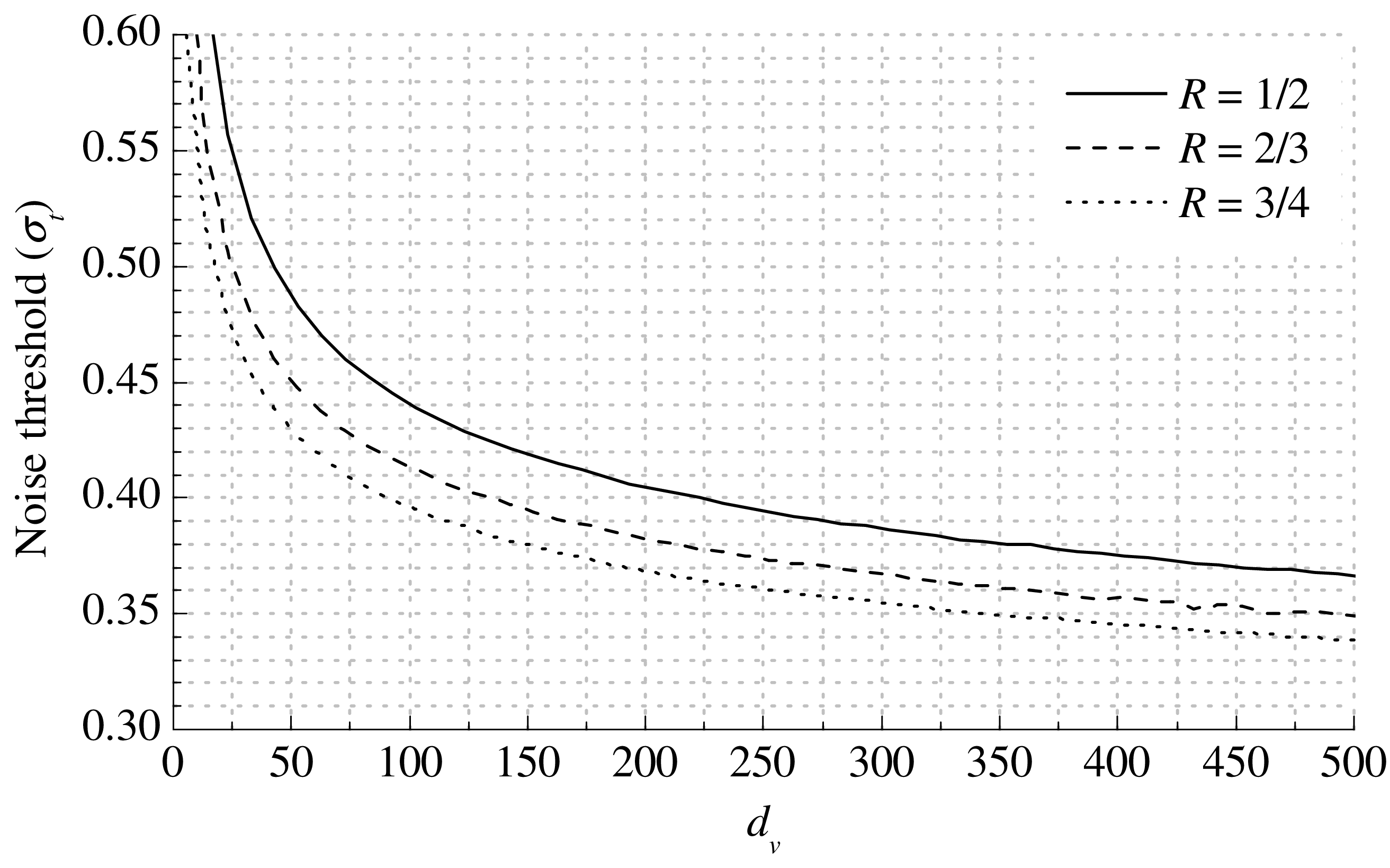}
\caption{Gaussian noise threshold values ($\sigma_t$) found through density evolution for regular codes with parity-check matrix column weight $d_v$ and rate $R = 1/2, 2/3, 3/4$.}
\label{fig:SigmaThresholds}
\par\end{centering}
\end{figure}

A first attempt that Eve can make is to also use an iterative soft-decision decoding algorithm to decode the public code.
The performance of these decoders is actually difficult to predict from a theoretical standpoint.
However, an ultimate bound on their performance can be computed through the density evolution technique \cite{Richardson2001}.
This provides the maximum noise level which can be compensated under the hypothesis of infinite length codes with
absence of closed loops in their associated graphs.
Obviously, when practical, finite length codes with cycles in their graphs are used, the maximum noise
level which still allows error correction is bounded away from the density evolution threshold.
However, the latter is still useful in our case, since it represents an ultimate limit.
Therefore, if the intentional noise level used during encryption is above the density evolution threshold,
we are sure that error correction cannot be performed through iterative soft-decision algorithms working
on the public code, independently of the code length.
The density evolution threshold decreases as long as the parity-check matrix column weight increases,
as shown in Fig. \ref{fig:SigmaThresholds}.
This is the reason why these decoding algorithms are very likely inefficient on the public code, which has a
dense parity-check matrix, unless a key recovery attack is first successfully accomplished.

However, even when the intentional noise level is above the density evolution threshold and iterative
soft-decision decoding algorithms are ineffective, an attacker could still exploit the soft reliability information 
to facilitate classical \acp{DA}.
For this purpose, the attacker could proceed as follows:
\begin{enumerate}
\item sort the ciphertext bits in decreasing reliability (\textit{i.e.}, $\left|\Lambda(x_i)\right|$) order,
\item select the $t_f$ least reliable ciphertext bits, suppose that they are in error and flip them,
\item use \ac{ISD} to correct the residual bit errors.
\end{enumerate}

The work factor of such an attack procedure can be computed as follows (mathematical derivations
are omitted for the sake of brevity).
The probability that the ciphertext bit at position $i$ in the ordered list is in error due to 
an intentional Gaussian noise with mean $0$ and standard deviation $\sigma$ can be computed as
\begin{equation}
P_{e,i}=n \binom{n-1}{i-1}\int_0^{+\infty}g_{-1,\sigma}(x) \left[P(x)\right]^{i-1} \left[1-P(x)\right]^{n-i}dx,
\label{eq:Pei}
\end{equation}
where $g_{-1,\sigma}(x) = \frac{1}{\sigma \sqrt{2 \pi}} e^{- \frac{(x+1)^2}{2 \sigma^2}}$ is the p.d.f. of a Gaussian variable with mean $-1$ and standard deviation $\sigma$ and
\begin{equation}
 P(x)=1-\frac{1}{2}\left[ \mathrm{erfc}\left(-\frac{x-1}{\sigma\sqrt{2}}\right)-\mathrm{erfc}\left(\frac{x+1}{\sigma\sqrt{2}}\right)\right].
\label{eq:Px}
\end{equation}
The probability that all the $t_f$ bits flipped by the attacker are indeed in error can then be computed as
\begin{equation}
P_{f} = \prod_{i=0}^{t_f-1} P_{e,n-i}.
\end{equation} 
The overall work factor of the decoding attack aided by the soft reliability information can be obtained as
\begin{equation}
WF_{\mathrm{DA}}^{(2)} = \frac{1}{\sqrt{r}} \frac{WF_{\mathrm{BJMM}}(n,k,t^*-t_f)}{P_{f}},
\label{eq:WFDA2}
\end{equation}
where $t^*$ is the total number of bit errors on the ciphertext induced by the intentional Gaussian noise.
Obviously, an attacker is free to choose the value of $t_f$ that minimizes the work factor expressed by \eqref{eq:WFDA2}.

If we use a purely random noise, the value of $t^*$ follows a binomial distribution with mean $\hat{t} = \frac{n}{2} \mathrm{erfc} \left(\frac{1}{\sigma \sqrt{2}}\right)$.
Therefore, some ciphertexts may experience small values of $t^*$, and hence may be more vulnerable to attacks of this type.
To avoid this risk, we can require Alice to discard those vectors $\mathbf{w}$ corresponding to values of $t^*$ falling below some threshold $\underline{t}$, and
compute the work factor considering the worst case in which $t^* = \underline{t}$.
We have verified through numerical simulations that a simple and effective choice is to impose that $t^* \ge \underline{t} = \hat{t}$.
This does not require Alice to perform too many attempts to generate the vector $\mathbf{w}$ (half of them are successful on average)
while allowing to achieve good security levels with compact keys.

\subsection{Key recovery attacks}
\label{subsec:KRA}

\acp{KRA} are aimed at recovering the private key from the public key.
Even when the private key is not exactly recovered, the attack may be successful by finding an
alternative private key which is still useful for an attacker to perform decoding.
An efficient way to recover a sparse parity-check matrix for the public code ($\mathbf{H'}$) is 
to search for its rows in the dual of the public code.
When $\mathbf{H'}$ is successfully recovered, it can then be used by an attacker to recover $\mathbf{H}$ 
by exploiting its sparsity, or to perform \ac{LDPC} decoding and correct the intentional errors.
In general, the matrix $\mathbf{H'}$ has column weight $d_v' = m \cdot d_v$ (that is, $d_v' = d_v$ for
the \ac{QC-MDPC} code-based system we consider) and row weight $d_c' = n_0 \cdot d_v'$.
Therefore, $d_v'$ must be large enough to make finding the rows of $\mathbf{H'}$ in the dual of the public 
code computationally infeasible for an attacker.

Since the codes are \ac{QC} with parity-check matrices in the form \eqref{eq:HCircRow}, all the rows 
of $\mathbf{H'}$ are obtained as the \ac{QC} shifts of one of them.
This reduces the attack complexity by a factor equal to the number of rows of $\mathbf{H'}$, \textit{i.e.}, $r$.
Therefore, the work factor of a \ac{KRA} can be computed as 
\begin{equation}
WF_{\mathrm{KRA}} = \frac{1}{r} WF_{\mathrm{BJMM}}(n,r,d_c').
\end{equation}

\section{Examples}
\label{sec:Examples}

Let us focus on $80$-bit security.
The McEliece cryptosystem with the shortest public key size known in the literature is
reported in \cite{Misoczki2012}, and achieves $4801$-bit public keys (with CCA2-security conversion)
using codes with length $n = 9602$, rate $1/2$ and a \ac{DFR} in the order of $10^{-7}$ or less.
Such a system uses $t = 84$ binary intentional errors and \ac{QC-MDPC} codes with public and private 
parity-check matrix column weight $d_v' = d_v = 45$. 

Let us consider a similar system with the same value of $d_v' = d_v$, and suppose that the same 
number of bit errors are induced by an intentional Gaussian noise, i.e., $\underline{t} = \hat{t} = 84$.
If we reduce the code length to $n = 7202$, we obtain $\sigma = \frac{1}{\sqrt{2} \mathrm{erfc}^{-1} \left(2 \hat{t}/n \right)} = 0.44091$.
We have verified through numerical simulations that, for these values of $n$ and $\sigma$, 
the \ac{LLR}-\ac{SPA} decoder is still able to compensate the intentional noise with a \ac{DFR} in the order of $10^{-7}$.

Concerning \acp{DA} and \acp{KRA}, their work factors are $WF_{\mathrm{DA}}^{(2)} = 2^{80.49}$ (minimum for $t_f=0$)
and $WF_{\mathrm{KRA}} = 2^{80.17}$, respectively.
According to Fig. \ref{fig:SigmaThresholds}, the density evolution threshold for codes with rate $1/2$ 
falls below $0.4$ for parity-check matrix column weights $> 223$.
Unless a \ac{KRA} is performed, the parity-check matrices of the public code which are available to an attacker are dense, 
with column weight in the order of one thousand or more.
Hence, iterative soft-decision decoders cannot be used to cancel the intentional noise.
Therefore, this system is able to achieve $80$-bit security with $3601$-bit public keys,
that is, about $25 \%$ less than the best known solution.

If we focus on $128$-bit security, the solution provided in \cite{Misoczki2012} that achieves 
the smallest public key is that using codes with rate $1/2$, $n = 19714$, $d_v' = d_v = 71$ and $t = 134$ intentional errors.
Considering $n=15770$ and an intentional Gaussian noise with $\underline{t} = \hat{t} = 134$ yields $\sigma = 0.41897$,
that is still above the density evolution threshold for dense matrices.
We have verified through simulations that a \ac{QC-MDPC} code with rate $1/2$, $n=15770$ and $d_v = 71$
is able to compensate such a noise level under \ac{LLR}-\ac{SPA} decoding, with a \ac{DFR} in the order of 
$10^{-7}$.
In this case, the attack work factors are $WF_{\mathrm{DA}}^{(2)} = 2^{129.06}$ (minimum for $t_f=0$)
and $WF_{\mathrm{KRA}} = 2^{130.24}$.
Therefore, $128$-bit security can be achieved with $7885$-bit public keys, that is, about $20 \%$ less than the best known solution
(reported in \cite{Misoczki2012} and requiring $9857$-bit public keys).

In both these cases, the smallest work factors are found when no bits are flipped by the attacker based on their reliabilities (\textit{i.e.}, $t_f=0$).
However, this situation changes when the code rate is larger than $1/2$, since the advantage an attacker gains by exploiting flipped bits becomes significant.
For example, if we consider a code with $n=10779$ and rate $2/3$, with $\underline{t} = \hat{t}=53$ (that implies $\sigma=0.38736$) and $t_f=0$, we obtain $WF_{DA}^{(2)}=2^{80.84}$.
Instead, if the attacker tries to flip the least reliable bits before performing \ac{ISD}, the attack work factor can be reduced to $WF_{DA}^{(2)}=2^{75.53}$ (mininum for $t_f=22$)
and the value of $\underline{t}$ must be consequently increased in order to achieve $80$-bit security.

\section{Conclusion}
\label{sec:Conclusion}

We have verified that using real-valued intentional noise during encryption can be highly beneficial in order to achieve
\ac{QC-MDPC} code-based cryptosystems with compact keys.
Our results show that public key size reductions up to $25 \%$ with respect to the best known solutions can be obtained.
Future investigations will involve the chance to jointly use binary and real-valued intentional errors, as
well as real-valued noise samples conveying part of the secret message.

\newcommand{\BIBdecl}{\setlength{\itemsep}{0.01\baselineskip}}
\bibliographystyle{IEEEtran}
\bibliography{Archive}

\end{document}